\documentstyle[11pt,aaspp4]{article}
\def\ee{\varepsilon}
\def\beq{\begin{equation}}
\def\eeq{\end{equation}}

\def\gsim{\;\rlap{\lower 2.5pt
 \hbox{$\sim$}}\raise 1.5pt\hbox{$>$}\;}
\def\lsim{\;\rlap{\lower 2.5pt
   \hbox{$\sim$}}\raise 1.5pt\hbox{$<$}\;}

\begin{document}
\title{Emission Spectra from Internal Shocks \\
in Gamma-Ray-Burst Sources}
\bigskip
\author{Ravi P. Pilla}
\smallskip
\affil{Department of Physics, Columbia University,
538 W 120th street, New York, 10027, NY;
ravi@cuphyb.phys.columbia.edu}
\medskip
\author{Abraham Loeb}
\smallskip
\affil{Astronomy Department, Harvard University, 60 Garden Street,
Cambridge, MA 02138; aloeb@cfa.harvard.edu}

\begin{abstract}
Unsteady activity of $\gamma$-ray burst sources leads to internal shocks in
their emergent relativistic wind.  We study the emission spectra from such
shocks, assuming that they produce a power-law distribution of relativistic
electrons and posses strong magnetic fields.  The synchrotron radiation
emitted by the accelerated electrons is Compton up-scattered multiple times
by the same electrons. A substantial component of the scattered photons
acquires high energies and produces $e^{+}e^{-}$ pairs.  The pairs transfer
back their kinetic energy to the radiation through Compton scattering.  The
generic spectral signature from pair creation and multiple Compton
scattering is highly sensitive to the radius at which the shock dissipation
takes place and to the Lorentz factor of the wind.  The entire emission
spectrum extends over a wide range of photon energies, from the optical
regime up to TeV energies. For reasonable values of the wind parameters,
the calculated spectrum is found to be in good agreement with the burst
spectra observed by BATSE.
\end{abstract}
\centerline{submitted to {\it ApJ Letters}, October 1997}

\section{Introduction}

The detection of FeII and MgII absorption lines at a redshift of $z=0.835$
in the optical spectrum of GB970508 (Metzger et al. 1997), provided the
first confirmation that $\gamma$-ray bursts (GRBs) originate at
cosmological distances.  Most of the qualitative properties of cosmological
GRBs are explained by the fireball model (see e.g., Goodman 1986;
Paczy\'nski 1986; M\'esz\'aros \& Rees 1993 [MR]).  In this model, a
compact ($\sim 10^{6-7}$ cm) source releases an energy of $\sim 10^{52}$
erg over a duration $t_{d}\la 10^2$ seconds with a negligible baryonic
contamination ($\la 10^{-5}M_\odot$).  Unsteady activity of the source
results in a wind composed of many thin layers (fireball shells) of varying
energy and baryonic mass. Within each shell the high energy-density at the
source results in an optically thick $e^{+}e^{-}$-pair plasma that expands
and accelerates to relativistic speeds. After an initial acceleration
phase, the radiation and thermal energy of the fireball plasma is converted
into the kinetic energy associated with the radial motion of the protons.
Collisions between the shells can convert part of that kinetic energy into
radiation and yield the primary GRB via synchrotron emission and
inverse-Compton scattering (Paczy\'nski \& Xu 1994; Rees \& M\'esz\'aros
1994 [RM]; Sari \& Piran 1997 [SP]).  As the wind continues to expand, it
impinges on the surrounding medium and eventually drives a relativistic
blastwave in it, which heats fresh gas and accelerates electrons to
relativistic speeds, thus producing the delayed afterglow radiation
observed on time scales of hours to months (van Paradijs et al.  1997; Bond
1997; Djorgovski et al.  1997; Mignoli et al.  1997; Frail et al.  1997)
via synchrotron emission (Wijers, Rees, \& M\'esz\'aros 1997;Waxman
1997a,b; Vietri 1997a,b) .

The primary GRB emission is more likely caused by internal shocks than the
external shocks (MR), since they can occur closer to the source and thus
account for the rapid variability observed in many bursts (RM; SP).
Unsteady activity of the central source naturally results in faster
shells overtaking slower ones in front of them, and hence in energy
dissipation by internal shocks. The complex temporal structure observed in
GRBs then reflects the activity-history of their sources (SP; Kobayashi,
Piran, \& Sari 1997 [KPS]).

It is often assumed that behind internal shocks, electrons are Fermi
accelerated with a near equipartition energy density and magnetic fields
acquire nearly equipartition strength.  The electrons cool by synchrotron
emission and inverse-Compton (IC) scattering off the synchrotron photons.
Under typical conditions, the time scale for IC scattering is shorter than
the synchrotron cooling time.  Multiple scattering of the photons boosts a
significant fraction of the radiation energy to frequencies above the
$e^{+}e^{-}$-pair creation threshold.  The pairs produced in this process
are also relativistic and cool rapidly by IC scattering. Since the
annihilation time of these pairs is longer than the hydrodynamic time in
the comoving frame, they survive in the wind for a long time. Although the
creation of pairs and their subsequent cooling is likely to leave
noticeable imprints on the emergent radiation spectrum, it has not been
analyzed before in the GRB literature.  Since the photon and electron
densities decline rapidly with radius, the strength of these signatures can
serve as a probe of the radius at which the internal shocks occur.

The GRB spectrum should also depend on the level of baryonic contamination
in the wind. The extreme limit of pure energy release with no baryons was
ruled out in the past based on the prediction that a point explosion of
this type would lead to a roughly thermal spectrum (Goodman 1986).  One can
place a lower limit on the baryonic mass in the fireball shells by
requiring that internal shocks should occur before an external shock
does (this limit depends on the ambient medium density).  On the other
hand, an upper limit can be placed based on the variability time scale of
the source and the condition that the shells be optically thin at the
radius where internal shocks occur.

In this {\it Letter} we study in detail the emergent spectra from the
collision of two fireball shells\footnote{In principle, each individual
peak in the burst light-curve might correspond to a collision of a pair of
shells (KPS). The time-averaged flux of the entire burst is then an
energy-weighted sum of the contributions from individual collisions. Here
we provide the time-averaged spectra for the collision of two shells,
whereas the generalization to the entire burst (with multiple peaks) is
straightforward.}.  In particular, we quantify the significance of the
radiation processes which were previously ignored in the literature, such
as multiple Compton scattering, ${\rm e^+e^-}$ creation, and their
subsequent cooling.  We use the collision kernels, reaction rates, and the
computational techniques given in Pilla \& Shaham (1997 [PS]). More details
about this calculation and an elaborate study of the spectral
characteristics of relativistic shocks will be included in a subsequent
publication (Pilla \& Loeb 1997).  In \S 2 we describe our model and
specify the physical conditions in the emission region.  In \S 3 we outline
the relevant radiation processes and compute the model spectra. Finally, \S 4
summarizes the main implications of this work.

\section{Physical Properties of Internal Shocks}

The typical fireball dynamics (Piran, Shemi, \& Narayan 1993, [PSN];
M\'esz\'aros, Laguna, \& Rees 1993; Sari, Narayan, \& Piran 1996 [SNP]; SP;
and KPS) can be illustrated by considering a single shell of total energy
$E=10^{51}E_{51}$ erg, rest mass $M = 10^{27}M_{27}$ g, and initial radius
$r_{0} = 10^{7}r_{7}$ cm. After a brief acceleration phase, the Lorentz
factor of the shell reaches a constant value $\Gamma$ at an observer-frame
radius $r_{m}\approx \Gamma r_{0}$ (the protons are taken to be
non-relativistic in the comoving frame, before the collision of shells).
The energy of the shell is predominantly kinetic beyond this stage.
Outside the radius $r\approx r_{exp}=2\Gamma^{2}r_{0}$ (all radii in the
present analysis are measured in the observer's frame), the comoving width
of the shell increases linearly with radius (PSN).  The comoving proton
density scales as $n\propto r^{-2}$ for shell radii $r\leq r_{exp}$ and as
$n\propto r^{-3}$ for $r>r_{exp}$. The Thomson optical depth of the shell
is $\tau\approx\left(r_{t}/r\right)^{2}$ for $r\ge r_{exp}$, where
$r_{t}\approx\left(M\sigma_{T}/4\pi m_{p}\right)^{1/2}\approx 6.3
\times 10^{12} M_{27}^{1/2}$ cm is the radius at which the shell becomes 
optically thin to
Thomson scattering by its own electrons. Here $\sigma_{T}$ is Thomson cross
section and $m_{p}$ is the proton rest mass. For $r>r_{exp}$, the comoving
width of the shell is $\delta\approx r/2\Gamma$.  Two photons which are
emitted with a proper-time difference $\sim \delta/c$, reach the observer
with a time separation $\sim \delta/2\Gamma c\approx r/4\Gamma^{2}c$ (SP).
Assuming that the radiative cooling time is much shorter than the light
transit time through the system, one finds that the observed width of the
radiation pulse $t_{p}$ and the radius of the emission region $r_{e}$ are
related through $r_{e}\approx 2\Gamma^{2}ct_{p}$, which in turn can be used
to constrain the values of $\delta, \tau$, and $n$ from observations.  The
emission spectrum will be nonthermal only if $r_{e}\gg r_{t}$.

Now consider a wind of total duration $t_{d}$, composed of many thin
fireball shells of thickness $\delta \ll ct_d$ ($\delta$ may change from one
shell to another).  We assume that for $r>r_{exp}$ there are regions in the
wind where slower shells precede the faster ones, i.e. $d\Gamma(r)/dr<0$,
where $\Gamma(r)$ is the local Lorentz factor. The spatial extent of the
wind is typically $\sim ct_{d}\gg r_{0}$ (in fact, $\delta\gg r_{0}$ also
holds in general). It was shown by Waxman \& Piran (1994) that under these
conditions the wind layers are susceptible to  Rayleigh-Taylor instability
because a rarefied fluid shell is pushing against a denser one. The
resulting turbulent mixing 
will complicate the shock structure and deform it away from a simple planar
geometry.  Merging of a rarefied shell with a denser one might therefore be
accompanied by the formation of ``fingers'' perpendicular to the shell
walls, similar to the non-relativistic shock structure in supernova
remnants (e.g., Jun \& Norman 1996).  The combined shells would then break
into bubbles of different sizes, and the energy dissipation would take
place near the bubble walls, due to collisions among them or instabilities
on their surfaces (Kamionkowski \& Freese 1992).

By assuming that the shells as well as the shock fronts remain planar, KPS
had found that the dissipation efficiency of internal shocks might obtain
high values $(\sim 50\%)$ for reasonable wind parameters.  We assume that
similar efficiencies are achieved in the case of unstable mixing.  Since
the combined area of the bubble walls greatly exceeds that of a planar
shock, the electron acceleration efficiency in the present case is likely
to be higher. However the temporal and spectral characteristics of the
bursts might be different in the two cases. For a planar shock the
accelerated electrons populate a thin layer around the shock front since
their cooling time is much shorter than the transit time of sound waves
across the shells.  However, in the unstable mixing case we assume, to a
first approximation, that the energy dissipation takes place nearly
uniformly and simultaneously throughout the entire volume of the emission
region.  For the purpose of estimating the physical conditions involved, we
take $E\approx 10^{51}$ erg, $M\approx 10^{27}$ g, $r_{e}\approx 10^{14}$
cm, and a bulk Lorentz factor of the emission region of $\Gamma\approx
400$.  These values yield $n\approx 3.1\times 10^{10}$ ${\rm cm}^{-3}$,
$\tau\approx 3\times 10^{-3}$, a comoving width of the post-shock shell
$\Delta\approx 1.4\times 10^{11}$ cm, and an average Lorentz factor of the
protons in the comoving frame (i.e., after shock heating) of
$\overline{\gamma}_{p}\approx 3$.

\section{Radiation Mechanisms and Spectra}

Acceleration of electrons to relativistic energies and the presence of
strong magnetic fields are essential for converting the energy dissipated
by the shock waves into radiation. Since the physics of neither of these
processes is well understood, we parameterize the corresponding energy
densities in units of their equipartition values.  A magnetic equipartition
parameter $\zeta_{B}=B^{2}/8\pi nm_{p}c^{2}(\overline{\gamma}_{p}-1)$,
corresponds to a field strength $B\approx
1.9[\zeta_{B}n_{10}(\overline {\gamma}_{p}-1)]^{1/2}10^{4}$ G,
where $n_{10}=n/(10^{10}{\rm cm}^{-3})$ (all quantities are comoving,
unless stated otherwise). We assume that the electrons are accelerated
throughout the emission region.  The amount of energy transferred from
protons to electrons is uncertain, and we define the acceleration
efficiency $\zeta_{e}$ in such a way that the average Lorentz factor of the
electrons immediately after they are heated (via Fermi-type acceleration)
is $\overline{\gamma}_{0} =
\zeta_{e}(\overline{\gamma}_{p}-1)m_{p}/m_{e}$, where $m_{e}$ is the
electron rest mass. Because the Coulomb collision time $\sim
1/cn\sigma_{T}\ln{\Lambda}\sim\Delta/c\tau\ln{\Lambda}\gg\Delta/c$, we can
safely ignore collisional relaxation in our analysis (here
$\ln{\Lambda}\sim 30$ is the Coulomb logarithm).

Collisionless acceleration of electrons can be efficient if they are
tightly coupled to the protons and the magnetic field by means of plasma
waves (Kirk 1994). The typical Alfv\'en speed in the plasma is
$\upsilon_{A}=B/(4\pi n m_{p})^{1/2} ={\rm min}(1,1.4\zeta_{B}^{1/2})c$.
The Larmor radius of an electron with a Lorentz factor $\gamma$ is
$r_{L,e}\approx 0.1 \gamma(\zeta_{B}n_{10})^{-1/2}$ cm, and that of a
proton of equal Lorentz factor, $r_{L,p}$, is larger by a factor
$m_{p}/m_{e}$.  The corresponding acceleration time scales (e.g., Hillas
1984) are, $t_{acc,e}\sim cr_{L,e}/\upsilon_{A}^{2}\simeq 1.7\times
10^{-13}\gamma(\zeta_{B}^{3}n_{10})^{-1/2}$ sec and $t_{acc,p}
\sim (m_{p}/m_{e})t_{acc,e}$, respectively. Synchrotron losses limit the 
maximum value of the electron Lorentz factor to $\gamma_{max,e}\sim
3.2\times 10^{5}(n_{10}^{3}\zeta_{B})^{-1/4}$. The minimum value is
determined by $\overline{\gamma}_{0}$ and the shape of the electron
distribution. We assume that the fraction of electrons per unit Lorentz
factor $\gamma$ has the form
$F_{e}(\gamma)=(p-1)\gamma_{min}^{p-1}\gamma^{-p}$ for
$\gamma_{min}\leq\gamma\leq\gamma_{max}$. Thus
$\gamma_{min}=(p-2)\overline{\gamma}/(p-1)$, where $\overline{\gamma}$ is
the average Lorentz factor at any given time. The energy density in
electrons immediately after their acceleration is
$u_{0}=nm_{e}c^{2}\overline{\gamma}_{0}$.  In our model the radiation time
scale is much shorter than the hydrodynamic expansion time scale in the
comoving frame (which leads to a radiative efficiency of nearly $100\%$)
and the radiation density at the end of electron cooling is therefore
$u_{0}$. We compute the spectra for $\zeta_{B}=0.1$, $\zeta_{e}=0.3$, and
$p=3.5$.

\parindent=0cm
{\it Electron Cooling and Radiation Spectrum}
\parindent=1cm

We assume that the radiation energy density is initially small, and 
hence the electrons start losing their energy via synchrotron emission.  As the
energy density of the emitted radiation builds up, cooling via IC
scattering becomes important as well.  The typical time scale for
synchrotron or IC losses is $t_{c}\approx q/\lambda cn\sigma_{T}\overline{
\gamma}^{2}$, where $\lambda=\zeta_{B}/\zeta_{e}$ in the synchrotron case and
 $\lambda=u_{\gamma}/u_{0}$ in the IC case.  Here $u_{\gamma}$ is the
radiation energy density and $q$ is a dimensionless constant of order
unity. For typical GRB conditions $t_{acc,e}\ll t_{c} \ll t_{0}$, where
$t_{0}\approx \Delta/c$ is the light transit time through the system.
Therefore electron cooling takes place only locally.  We assume that the
acceleration proceeds throughout the cooling phase in such a way that it
maintains a steady power-law distribution with a constant index $p$, while
$\overline{\gamma}$ declines due to radiative losses.
%
If $\zeta_{B}\ll\zeta_{e}$, the synchrotron cooling time is long and
multiple IC scatterings become important.  Each scattering in the Thomson
regime increases the photon energy by a factor
$4\langle\gamma^{2}\rangle/3$ (Loeb, McKee, \& Lahav 1991), so that some of
the photons are eventually boosted into the Klein-Nishina (KN) regime. As
$\overline{\gamma}$ decreases, the up-scattered part of the spectrum
spreads over many decades in frequency.  The electron energy density
$u_{e}=nm_{e}c^{2}\overline{\gamma}$ changes at a rate
\beq
\frac{du_{e}}{dt}=\left(\frac{du_{e}}{dt}\right)_{sy}+
\left(\frac{du_{e}}{dt}\right)_{IC},
\label{energy-rate}
\eeq
where $\left(du_{e}/dt\right)_{sy}=-4u_{B}/3T$, $u_{B}=B^{2}/8\pi$,
$1/T=cn\sigma_{T}\langle\gamma^{2}\rangle_{0}$, and
$\langle\gamma^{2}\rangle_{0}$ is the average value of $\gamma^{2}$ over
the initial electron distribution. The IC cooling rate,
$\left(du_{e}/dt\right)_{IC}$, is derived below. The characteristic
cooling time $t_{c}$ is a few times $T$ under typical conditions.  If
$u_{\gamma}(t)$ is the instantaneous energy density of radiation, then
energy conservation implies $u_{\gamma}(t)+u_{e}(t)=u_{0}$ (which is constant),
and by
assumption $u_{\gamma}(0)=0$.  The instantaneous radiation spectrum can be
characterized by $\Phi(\ee,t)$ so that
$u_{\gamma}(t)\Phi(\ee,t)d\ee=\upsilon(\ee,t)d\ee$ is the fraction of the
radiation energy density within an interval $d\ee$ around $\ee$; here
$\ee=h\nu/ m_{e}c^{2}$, $\nu$ is the photon frequency, and $h$ is Planck's
constant.

The spectral evolution rate (see Pilla \& Loeb 1997) 
is derived from the equation
\beq
u_{\gamma}(t)\frac{\partial}{\partial t}\Phi(\ee,t)-
\Phi(\ee,t)\frac{du_{e}}{dt}=\left[\frac{\partial}{\partial t}
\upsilon(\ee,t)\right]_{sy}+ \left[\frac{\partial}{\partial t}
\upsilon(\ee,t)\right]_{IC},
\label{spec-rate}
\eeq
where we have used the fact that $du_{\gamma}/dt=-du_{e}/dt$. The first
term on the right hand side is the synchrotron emissivity from relativistic
electrons (Rybicki \& Lightman 1979) which depends on $F_{e}(\gamma,t)$;
the second term is 
\beq
\left[\frac{\partial}{\partial t}\upsilon(\ee,t)\right]_{IC}=
\frac{u_{\gamma}}{2T}\int_{-1}^{1}d\mu(1-\mu)\int_{\gamma_{l}}^{\gamma_{max}}
d\gamma F_{e}(\gamma,t)\frac{\Phi(\ee_{1},t)}{\langle\gamma^{2}\rangle_{0}}
\zeta\frac{\sigma_{KN}}{\sigma_{T}},
\label{eq:dvdt}
\eeq
where $\mu$ is the cosine of the scattering angle, $\gamma_{l}=max
(\gamma_{min},\ee)$, $\zeta=\gamma/(\gamma-\ee)=1+2\gamma(1-\mu)\ee_{1}$,
$\ee_{1}=\ee/2\gamma(1-\mu)(\gamma-\ee)$, and $\sigma_{KN}=3\sigma_{T}
(\zeta^{2}-2\zeta/3+1)/4\zeta^{3}$ is the Klein-Nishina cross-section. 
By integrating both sides of equation~(\ref{eq:dvdt}) over all values of
$\ee$, we obtain $\left(du_{\gamma}/dt\right)_{IC}$
on the left hand side, whereas on the right hand side we use the relation
$\left|d\ee/d\ee_{1}\right|= 2(1-\mu)\gamma^{2}/\zeta^{2}$ to convert the
integral to be over $\ee_{1}$ and integrate over all values of $\gamma,
\ee_{1}$, and $\mu$.  Finally, we divide the result by 2 to avoid double
counting of each scattering event, and obtain
\beq
\left(\frac{du_{\gamma}}{dt}\right)_{IC}=-\left(\frac{du_{e}}{dt}\right)_{IC}
=\frac{u_{\gamma}}{2T}\int_{-1}^{1}d\mu\int_{\gamma_{min}}^{\gamma_{max}}
d\gamma F_{e}(\gamma,t)\int_{\ee_{min}}^{\ee_{max}}d\ee_{1}\Phi(\ee_{1},t)
\frac{\gamma^{2}}{\langle\gamma^{2}\rangle_{0}}(1-\mu)^{2}\frac{1}{\zeta}
\frac{\sigma_{KN}}{\sigma_{T}},
\label{IC-rate}
\eeq
where $\ee_{min,max}$ are the limiting energies of photons in
the plasma. In the Thomson regime $\zeta=1$ and $\sigma_{KN}=\sigma_{T}$,
and one finds
\beq
\left(\frac{du_{\gamma}}{dt}\right)
_{IC}=\frac{4u_{\gamma}\langle\gamma^{2}\rangle}
{3T\langle\gamma^{2}\rangle_{0}}=\frac{4}{3}cn_{e}\sigma_{T}u_{\gamma}
\langle\gamma^{2}\rangle,
\eeq
in agreement with the well-known result for $\gamma\gg 1$ (Loeb et al.
1991).  The coupled equations (\ref{energy-rate})--(\ref{IC-rate}) are
solved numerically for the radiation spectrum at the end of the cooling
process (i.e., when $u_{\gamma}\rightarrow u_{0}$). Cooling ends after a
relatively short time, $\la 10 T \ll t_{0}$, for the parameters of interest
here, and the photons decouple from the electrons subsequently. An example
for the time evolution of the spectrum is shown in Figure 1a. It is evident
that the radiation density above the $e^{+}e^{-}$-pair creation threshold
is substantial.

The spectral evolution rate due to pair creation is described by
\begin{eqnarray}
\frac{\partial}{\partial t}\left[\upsilon(\ee,t)\right]_{\pm}&=&
u_{\gamma}(t)\frac{\partial}{\partial t}\left[\Phi(\ee,t)\right]_{\pm}
+\Phi(\ee,t)\left(\frac{du_{\gamma}}{dt}\right)_{\pm}\nonumber\\
&&\nonumber\\
&=&\frac{u_{\gamma}}{T^{\prime}}{\cal I}_{1}(\ee,t)\equiv 
-\frac{u_{\gamma}}{4T^{\prime}}
\int d\mu \int d\ee^{\prime} \frac{\Phi(\ee,t)}{\ee}\frac{\Phi(\ee^{\prime},t)}
{\ee^{\prime}}\frac{\ee}{\overline{\gamma}_{0}}(1-\mu)\frac{\sigma_{\pm}}
{\sigma_{T}},
\label{eq:pairc}
\end{eqnarray}
where the integration is over the range $-1\leq \mu \leq 1$ and
$\ee_{min}\leq\ee\leq\ee_{max}$ subject to the condition $\ee\ee^{\prime}
(1-\mu)>2$. Here $\sigma_{\pm}$ is the pair creation cross section (PS) and
$1/T^{\prime}=cn\sigma_{T}\overline{\gamma}^{2}_{0}$. In all examples
considered in this {\it Letter}, we find that $t_{c}{\cal I}_{1}
(\ee,t)/T^{\prime}\ll 1$ for all relevant photon energies. Therefore, pair
creation comes into play after the original electrons cool and decouple.
The total energy loss per unit time and volume due to pair creation is
obtained by integrating equation~(\ref{eq:pairc}) over all values of $\ee$,
\beq
\left(\frac{du_{\gamma}}{dt}\right)_{\pm}=\frac{u_{\gamma}}{T^{\prime}}
\int_{\ee_{min}}^{\ee_{max}}d\ee {\cal I}_{1}(\ee,t)\equiv
\frac{u_{\gamma}}{T^{\prime}}\widetilde{\cal I}_{1}(t).
\eeq
The rate at which pairs are created per unit volume is
\beq
\left(\frac{dn}{dt}\right)_{\pm}=\frac{n}{T^{\prime}}\int_{\ee_{min}}^
{\ee_{max}
}d\ee {\cal I}_{2}(\ee,t)\equiv
\frac{n}{T^{\prime}}\widetilde{\cal I}_{2}(t),
\eeq
where ${\cal I}_{2}(\ee,t)$ is same as ${\cal I}_{1}(\ee,t)$ except that
$\ee/\overline{\gamma}_{0}$ is absent in its integrand. At the beginning of
pair creation, the energy density of radiation is $u_{0}$. The average
Lorentz factor of the newly created pairs is $\overline{\gamma}_{\pm}(t)
\approx\overline{\gamma}_{0}\widetilde{\cal I}_{1}(t)
/\widetilde{\cal I}_{2}(t)$,
since $u_{\gamma}/n\approx \overline{\gamma}_{0}$. Because the pairs are
born relativistic, they transfer almost all their energy to the radiation
via IC scattering (and also synchrotron emission if $\zeta_{B}\approx
\zeta_{e}$) on a time scale of order a few $T^{\prime}$.
Similar cascades of pair creation and cooling occur in active galactic
nuclei (Svensson 1987).  Throughout the pair cascade and cooling process
the radiation spectrum evolves continuously but its total energy density
remains nearly constant at a value close to $u_{0}$.  Figure 1b shows an
example of this evolution up to a time $\approx t_{0}$. Because the
hydrodynamic time scale in the comoving frame $t_{hyd}\approx
r/c\Gamma\approx 2t_{0}$, pair processes can not continue to operate for
times much longer than $t_{0}$ because of the decline in the densities due
to the expansion of the fireball.

\parindent=0cm
{\it Spectral Predictions of the Model}
\parindent=1cm

We solved the coupled equations given above using the methods described by
PS and obtained the model spectra for $E\approx 1.1\times 10^{51}$ erg,
$\overline{\gamma}_{p}=3$, $\zeta_{B}=0.1$, $\zeta_{e}=0.3$, and $p=3.5$.
We assume a source redshift $z_s\approx 1$ and include its effect on photon
energy and flux.  The total observable fluence for the above parameters is
$\approx 3.1\times10^{-7}\,\,
\mbox{erg}/\mbox{cm}^{2}$.

Figure 1 shows the entire spectral evolution (in the comoving frame) due to
the synchrotron and IC cooling and pair cascade processes. We take
$M\approx 10^{27}$ g which yields $\Gamma\approx
E/\overline{\gamma}_{p}Mc^{2}\approx 400$ (
which are consistent with empirical
constraints, e.g.  Woods \& Loeb 1995); and a dissipation (or shell
collision) radius of $3.3\times 10^{13}$ cm. Note that the assumed values
for $E$ and $M$ correspond only to a single emitting shell; they should
obviously be higher for the entire burst. For the above parameters, we
get $\Delta\approx 4.1\times 10^{10}$ cm, $n\approx 1.1\times
10^{12}\,\,\mbox{\rm cm}^{-3}$, $\tau\approx 3\times 10^{-2}$, and
$B\approx 6.3\times 10^{4}$ G. The characteristic time scale for
synchrotron and IC cooling is $T=1.6\times 10^{-5}t_{0}$. We find that
after a time $t=t_{c}\approx 16T$, only $10\%$ of the initial energy
remains with the electrons.  Figure 1a illustrates that as a result of
electron cooling, the location of the synchrotron peak shifts to lower
energies and also becomes broader.  Later on, another peak develops at a
much higher energy due to IC losses, with KN suppression in the very
high-energy tail.  Figure 1b shows the evolution much later due to the
pair-cascade process, which depletes the population of high-energy photons.
The IC loss by these pairs produces a power law tail at yet higher
energies.  The low-energy synchrotron part of the spectrum is almost
unaltered throughout this phase. The computation was stopped at $t\approx
t_{0}$ at which time  most of the photons leave the system.

Figure 2 shows the dependence of the emergent spectrum on the dissipation
radius for a fixed value of $\Gamma$ (panel [a]) and dependence on $\Gamma$
for a fixed value of the dissipation radius (panel [b]).  Panel (a) shows
that the location of synchrotron peak moves up in energy as the radius
decreases (with an opposite trend for the IC peak); hence the overall
extent of the emission spectrum gets narrower at smaller radii. Since
$B\propto n^{1/2}\propto r^{-3/2}$ and $\tau\propto r^{-2}$, 
as $r$ decreases the
synchrotron frequency increases and the pair-production depletion of
high-energy photons is enhanced.  One of the model spectra
in Figure (2a) is compared with the BATSE data through the empirical
formula for the time-integrated flux given by Band et al. (1993 [B93]). 
In that case our model predicts
$\sim 50\%$ more time-integrated flux at 10 keV than observed, which could
be due to an oversimplified form of the electron distribution function which we
have adopted.  
In our notation, the number of photons per unit interval of $\ee$ is $dN/d\ee
\propto \ee^{-1}\Phi(\ee)$. Therefore, Band's formula (B93) reads 
\begin{eqnarray}
\ee^{-1}\Phi(\ee)&=&c_{1}\ee^{-\alpha}\exp(-\ee/\ee_{0}),\qquad
\mbox{\rm if\,\,} \ee\leq(\beta-\alpha)\ee_{0}\nonumber\\
&&\nonumber\\[-10pt]
&=&c_{2}\ee^{-\beta},\qquad\mbox{\rm otherwise.}
\label{eq:band}
\end{eqnarray}
The constants $c_{1,2}$ are fixed by the requirement that this function be
continuous at $\ee=(\beta-\alpha)\ee_{0}$ and the normalization
$\int_{0}^{1}d\ee\Phi(\ee)=1$.  Note that we have altered the signs of both
indices $\alpha$ and $\beta$ relative to the convention of B93.  The
observed values of the parameters (cf. Table 4 in B93) are in the range
$0.3<\alpha<1.5$ (although in a few cases $\alpha$ is zero or negative),
$1.6<\beta<5$, and $15<\ee_{0}m_{e}c^{2}\,$(keV)$\,<3\times 10^{3}$.  The
majority of the bursts seem to have $\alpha\approx 1$, $\beta\approx 2$,
and $\ee_{0}m_{e}c^{2}\approx$ a few hundred keV.  From the model spectra
in Figure 2, it is clear that the emission extends over a wide range of
photon energies, from the optical to the TeV regime. 
We expect more optical emission when the dissipation takes place
at larger radii. It is very likely that internal shocks occur over a wide
range of radii, thereby extending the spectrum to longer wavelengths, with
simultaneous optical and x-ray emission during a GRB. Recent
reports of nearly simultaneous 
detection of x-ray emission from GB960720 (Piro et al. 1997),
GB970815 (Smith et al.  1997), and GB970828 (Remillard et al. 1997) are in
qualitative agreement with our expectation for internal shocks.

\section{Conclusions}

We have shown that the emission spectra from internal shocks are affected
by synchrotron emission, multiple Compton scatterings, and pair creation
(Fig.  1). Our model spectra could mimic the observed BATSE spectra (B93)
for reasonable choices of the shell Lorentz factor $\gamma\sim
10^2$--$10^3$ and shock radius $r\sim 10^{13}$--$10^{14}~{\rm cm}$ (Fig.
2).

The depletion of high energy photons due to pair creation is a sensitive
probe of both $\Gamma$ and $r$.  Detection of the high energy trough
present in Figures (1b) and (2) can therefore be used to constrain these
parameters.
The potential degeneracy between the values of $\Gamma$ and $r$ can be removed
on the basis of variability data, as the characteristic variability
time scale depends on a different combination of these parameters, $\sim
r/2\Gamma^2$.  Despite their high-energy cut-off, our model spectra extend
all the way up to photon energies as high as 10 GeV--TeV.

\acknowledgements
We acknowledge discussions with M. Kamionkowski, T. Piran, M. A.
Ruderman, and E. Woods. One of us (RP) thanks R. Sari for many critical
comments and stimulating discussions during the VIII Marcel
Grossmann meeting. This research was supported in part by NASA grants 
NAG5-618 and -2859 (for RP) and NASA ATP grant NAG 5-3085 and the Harvard
Milton fund (for AL).


\clearpage
\newpage
\begin{figure}[b]
\vspace{2.6cm}
\includegraphics{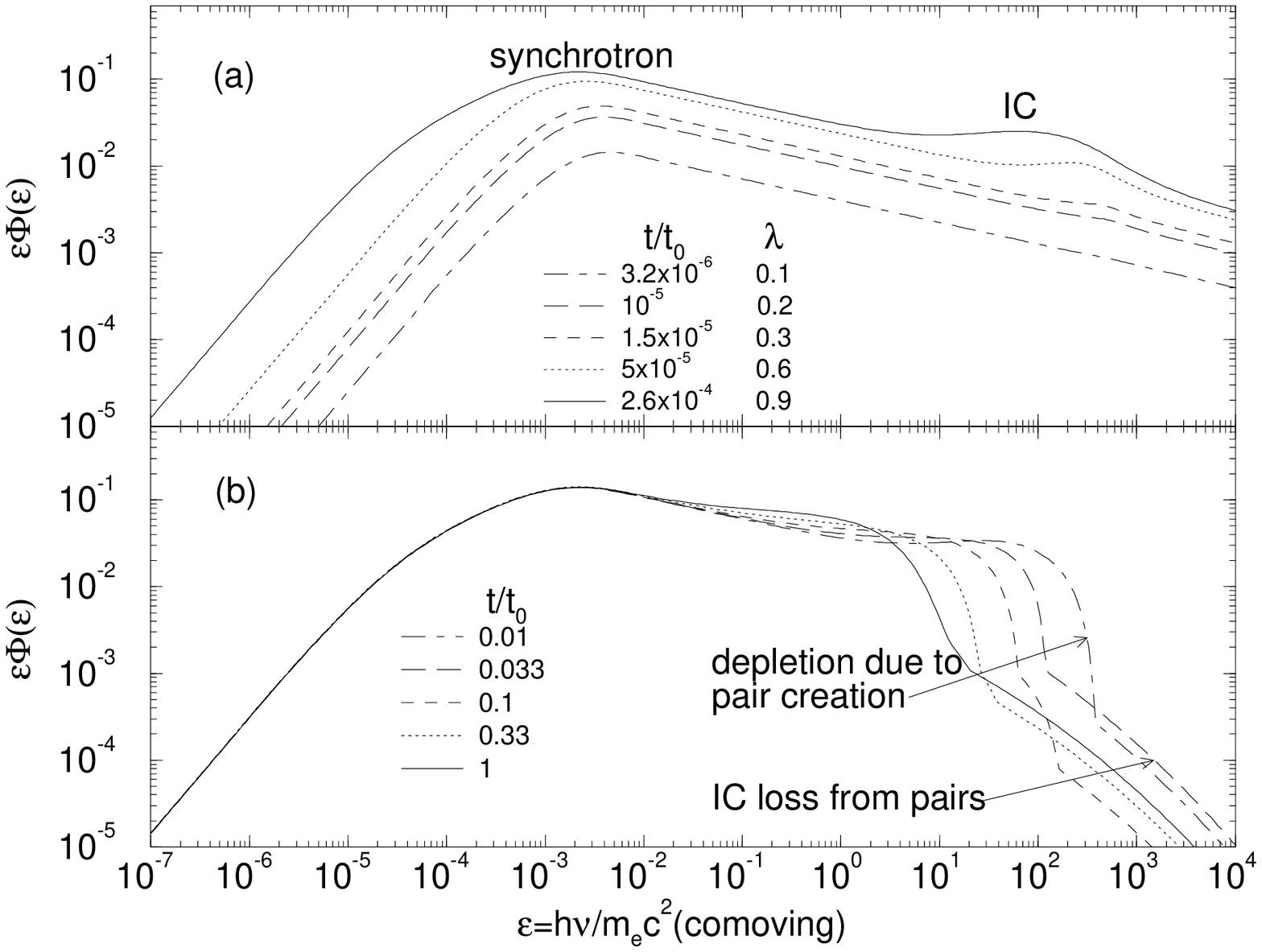}
\vspace*{3.5 in}
\caption[One] {Time evolution of the radiation spectrum for $r=3.3\times
10^{13}$ cm; the values of all other parameters are specified in the text.
Panel (a) shows the spectra at different times in the early phase of
electron cooling where only synchrotron and IC cooling are effective. The
normalized time and radiation density $\lambda=u_{\gamma}(t)/u_{0}$ are
indicated for each curve. Because $\zeta_{B}/\zeta_{e}=0.33$, the IC
component begins to grow only when $\lambda$ exceeds $\sim 0.3$, as
expected from equation (\ref{energy-rate}). Panel (b) shows the evolution
much later due to pair creation. The final density of (cold) electrons and
positrons is $\tilde{n}\approx 40n$ in this example.  However, they carry
only a negligible fraction ($\sim$ few percent) of the total energy.}
\end{figure}
\clearpage
\newpage
\begin{figure}[b]
\vspace{2.6cm}
\includegraphics{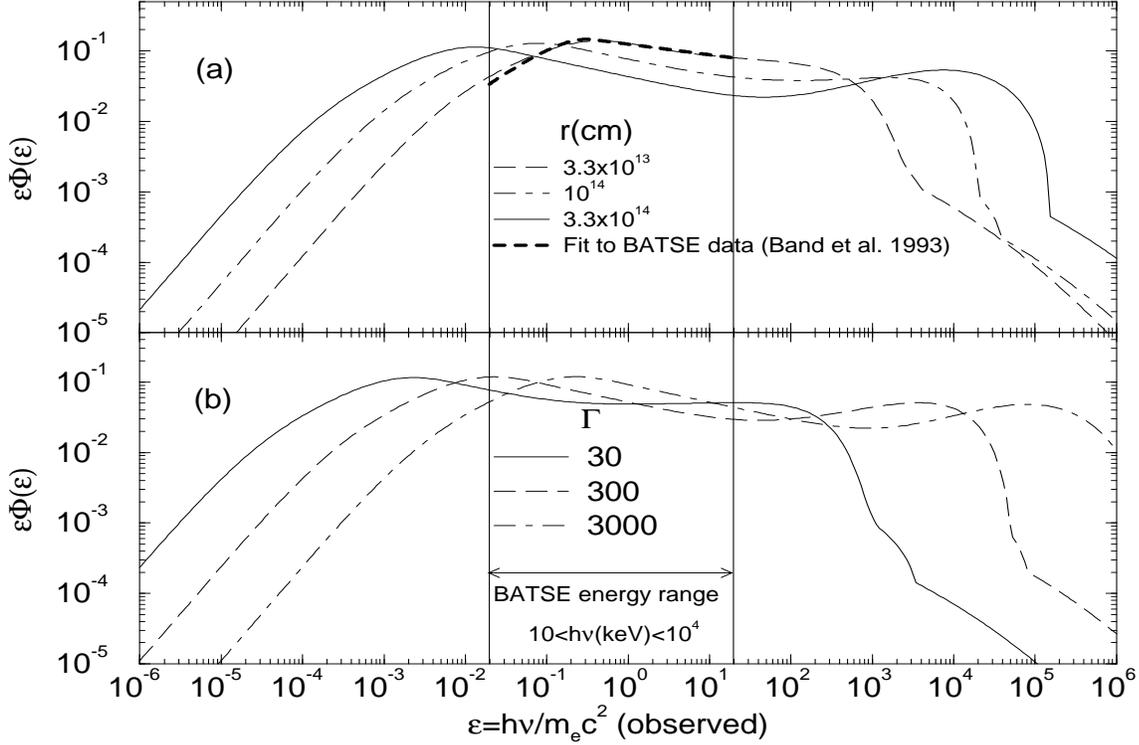}
\vspace*{3.5 in}
\caption[Two] {Dependence of the model spectra on the dissipation 
radius [panel (a)] and bulk Lorentz factor [panel (b)] of the emitting
shell. On panel (a) we show also the Band spectrum [cf.
Eq.~(\ref{eq:band})] for $\alpha=1.1$, $\beta=2.15$, and
$\ee_{0}m_{e}c^{2}=200$ keV, which is in a good qualitative agreement with
the predicted spectrum for $r=3.3\times 10^{13}$ cm.  The vertical lines
correspond to 10 keV and 10 MeV, which roughly bracket the BATSE energy
range.  On panel (b) we show the $\Gamma$-dependence of the spectra for a
fixed $r=2\times 10^{14}~{\rm cm}$. The inferred emission radii and Lorentz
factors are consistent with the empirical constraints set by the BATSE
variability data (e.g., Woods \& Loeb 1995).}
\end{figure}
\end{document}